# Rapidly Deployable Intelligent 5G Aerial Neutral Host Networks: an O-RAN-Based Approach

Yi Chu, David Grace, Josh Shackleton, Andy White, David Hunter and Hamed Ahmadi

*Abstract*—Rapidly deployable mobile networks are in demand to support connectivity in underserved areas and on occasions where the existing infrastructure becomes unavailable. Aerial platforms are ideal for delivering such networks benefitting from their flexibility and large coverage. However, compared with the terrestrial networks (TNs), the non-terrestrial networks (NTNs) have significantly different features in terms of energy consumption, network capacity and hardware design. This paper adapts the framework of Open Radio Access Network (O-RAN) and presents the potential of including NTNs into the O-RAN architecture. This gives the NTNs the access to the O-RAN's core intelligence, the RAN Intelligent Controller (RIC) which allows the optimisation of the network against various targets. This paper also presents a compact neutral host 5G network testbed using software defined radio (SDR) and open-source software. The small form factor of such networks allows them to operate with ground and aerial vehicles, therefore providing coverage at flexible locations without relying on existing infrastructure.

*Index Terms*— Open-source 5G, Open RAN, Helikite, rapid deployment.

## I. INTRODUCTION

Today, being able to receive consistent mobile network services is considered as a daily essential. According to the Connected Nations 2022 UK Report [1], 99% of premises receive outdoor coverage provided by at least one of the four UK mobile network operators (MNOs). However, the coverage of the UK landmass ranges from 80% to 87% across all MNOs. Mobile coverage inside vehicles on primary roads ranges from 83% to 88% but falls to 73% to 79% for secondary roads (which are likely to be in rural areas). These result in the landmass in remote areas of Scotland, UK National Parks and areas of outstanding natural beauty being underserved, with poor-quality (or no) Internet, and in many cases, no mobile coverage. The situation becomes worse in low-and middle-income countries where the people living in rural areas are 33% less likely to be connected to mobile networks compare with the people living in urban areas [2]. This occurs for two main reasons: the costs of installation and maintenance of the network in remote areas and regulatory restrictions. The rollout of 6G will occur additional capital costs for the MNOs to update hardware and install new terrestrial sites, therefore creating additional barriers for rural coverage. A rapidly deployable mobile network providing coverage in underserved areas can support temporary events (e.g., weddings and tourist events) and enables infrastructure dependent Internet of Things (IoT) applications such as connected autonomous vehicles (CAVs) and smart farming.

The flexibility and large coverage of aerial platforms make them ideal for delivering network infrastructure in rural areas. The Low Altitude Platforms (LAPs) such as drones and tethered balloons (e.g., Helikite [3]) have the features of low cost and flexible deployment, and the High Altitude Platforms (HAPs) have the advantages of large coverage and long endurance. These aerial platforms have been identified by 3GPP [4] as options for filling coverage gaps of TNs and attracted many discussions [5] as well as implementations [6]. However, compared to TNs, the NTN design needs to tackle additional challenges such as signal propagation with Doppler components, co-existence with TNs, user equipment (UE) mobility management, energy efficiency, cell coverage management and antenna design.

The future 5G and 6G networks are transforming into a flexible and disaggregated architecture to support the increasing connectivity demands and rapid deployments. Open Radio Access Network (O-RAN) is the enabling technology which decomposes the RAN, allowing flexible deployment and interoperability of the RAN components from multiple vendors, and most importantly, incorporating Artificial Intelligence (AI) into the network [7]. The RAN Intelligent Controller (RIC) [8,9] and the customisable apps it hosts enable intelligent control of the RAN in different timescales, with specific optimisation targets such as throughput, energy efficiency and co-existence. The relationship between O-RAN and Unmanned Aerial Vehicle (UAV) networks has drawn attention from the research community [10,11], and a testbed is in demand to evaluate the evolving theory. In this paper, we present a heterogeneous network architecture which incorporates TNs and multiple layers of NTNs, with the network performance optimised by the RICs located at the edge and cloud of the network. The aerial segment of such heterogeneous networks is still nascent, and a testbed is needed to identify issues during implementation which may form part of the thinking of O-RAN design for NTN. Therefore, we design a compact 5G neutral host network testbed which can be carried by ground and aerial

The research was supported by the project Mobile Access North Yorkshire, funded by the Department for Digital, Culture, Media & Sport (DCMS), and 5G Testbed and Trials programme, as well as the project Yorkshire Open Ran (YO-RAN), funded by the Department for Science, Innovation and Technology (DSIT), and the Future Open Networks Research Challenge. The authors Yi Chu (corresponding author), David Grace, Josh Shackleton, Andy White, David Hunter and Hamed Ahmadi are with the School of Physics, Engineering and Technology, University of York, York, United Kingdom (emails: yi.chu@york.ac.uk; david.grace@york.ac.uk; josh.shackleton@york.ac.uk; andy.white@york.ac.uk; david.hunter@york.ac.uk; hamed.ahmadi@york.ac.uk).

vehicles and conduct extensive experiments to evaluate the network.

The rest of this article is organised as follows. Section II provides an overview of the proposed heterogeneous network architecture. Section III introduces the design details of the aerial network testbed. Section IV presents the implemented testbed. Section V provides the evaluation of the testbed. Section VI discusses the challenges and the future pathways. Section VII concludes this article.

## II. Heterogeneous Network Architecture

Fig. 1 shows the proposed heterogeneous network architecture including three layers at different altitudes:
- Terrestrial layer: includes typical TN infrastructure and network users such as mobile handsets, CAVs and IoT devices. This layer also includes the terrestrial components of higher layers (low airborne layer in particular).
- Airborne layer: aerial platforms at altitudes below 30 km which can be further divided into:
  - Low airborne layer: LAPs such as drones and tethered balloons. These low-cost platforms can be rapidly deployed and will be the main focus of this paper. Tethered balloons (e.g., Helikite) can be used as temporary or semi-permanent NTN infrastructure, benefit from their long flight time and large payload capacity. Lightweight drones offer superior network mobility and enable various IoT applications.
  - High airborne layer: typical non-solar-powered aircraft and solar-powered stratospheric aircrafts such as the Sceye airship and Airbus Zephyr [12]. A single quasi-stationary HAP can cover an area of at least 60 km diameter and provide wireless access for ground users or backhaul for lower altitude NTNs. Depending on their size and payload capacity, massive multiple-input multiple-output (mMIMO) can be used to further improve the spectrum efficiency and capacity density. A comprehensive overview of the HAP technology can be found in [12].
- Orbital layer: a fleet of Low Earth Orbit (LEO) and geostationary (GEO) satellites which provide continuous global coverage. These permanent NTN infrastructures are suitable for supporting the backhaul for other NTNs, however, the unavoidable latency makes them difficult for carrying latency-sensitive services (e.g., fronthaul).

Different types of O-RAN components can be flexibly deployed across the three layers benefitting from the Virtual Network Functions (VNFs) and open interfaces. Frequency Range 1 (FR1) O-RAN Radio Units (O-RUs) can be deployed on the airborne layers to directly serve ground UEs or lower layer NTNs. Frequency Range 2 (FR2) O-RUs can be deployed on the high airborne layer and layers above to serve as xhaul for other NTNs. The logical components including the O-RAN Distributed Units (O-DUs), O-RAN Central Units (O-CUs), Near-Realtime RIC (Near-RT RIC [9]) and more speculatively a Realtime RIC (RT-RIC [10]) can be deployed together with the O-RU, at the local site edge close to the O-RU or at a cloud edge connected to the O-RU via a reliable fronthaul. The less latency-sensitive components such as the Non-Realtime RIC (Non-RT RIC [8]), Service Management and Orchestration (SMO) and Core Network (CN) can be deployed at the cloud. The open interfaces of the O-RAN architecture allow the operator to integrate networks with massively different capabilities from multiple vendors (e.g., small cell RUs on drones and mMIMO RUs on HAPs) into the same NTN network rather than having to deploy several separated networks of different types. The RIC and the E2 interface also make joint optimisation possible for multilayer NTNs. A more detailed deployment design for LAPs will be presented in Section IV.

With the AI/Machine Learning (ML) support of the RIC [13], the O-RAN research community has explored opportunities for optimisation. Below we present an indicative overview of optimisation directions.
- Energy saving: energy consumption determines the lifetime of most NTNs carried by aerial platforms as well as battery powered UEs. The RIC allows finer controls of the RAN such as RU on/off, RF power control and resource allocation which have been covered by many literatures for applying O-RAN to TNs.
- Traffic steering: active UE handover (HO) decisions can be made by xApps at a near-RT timescale for tasks such as cell load balancing and UE Quality of Service (QoS) optimisation. The traffic steering xApps collect UE/cell Key Performance Measurement (KPM) data and make HO decisions together with the policies sent from energy saving/route planning rApps. The mobile UEs requesting low-latency links (e.g., self-driving CAVs) can be moved to macro cells (e.g., HAPs) with tracking beams for mobility support and the UEs requesting high bandwidth (e.g., 4K video streaming) can be distributed across capacity boosting small cells (e.g., LAPs) to optimise the total throughput.
- Flight management for mobile platforms: this includes aircraft route planning, which serves the purpose of cell mobility/coverage management, and aircraft attitude control, which affects cell beam pointing and energy harvesting.
  - Route planning: The Non-RT RIC collects long timescale data such as UE types, locations/mobility and traffic patterns, and determines the number/type of NTNs to deploy as well as their flight paths according to the resources available. The route planning includes short term routes (e.g., the path for the drones to collect data from IoT devices) and long term paths (e.g., the trajectory for one HAP to serve a certain area) where different network states and optimisation targets are applied. Cloud-based coverage planning tools can be used to support the route planning with accurate coverage mapping.
  - Attitude control: unlike TNs with fixed antenna mounting, the attitudes of the aircraft (pitch, roll and yaw angles) affect the cell locations particularly on aircrafts equipped with directional antennas. Although beamforming can be used to track the cell locations, the beam shapes may change and produce unnecessary interference (e.g., grating lobes) with certain attitudes. LAPs with rapidly changing attitudes increase the difficulty of beamforming significantly. The aircraft attitudes also determine the orientation of the solar cells against the sun, which affects the energy harvesting efficiency. These factors should all be considered by the flight management.

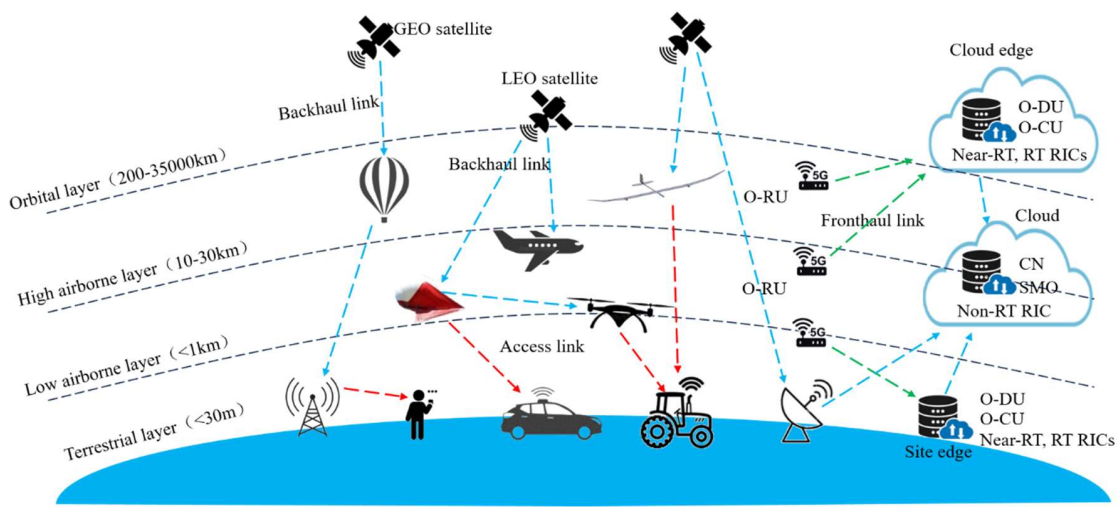

Fig. 1. Heterogeneous network architecture

## III. AERIAL NETWORK TESTBED DESIGN

The low airborne layer networks have the potential to become the most common coverage filler given their low-cost, flexible and ease of implementation nature. Fig. 2 shows a 5G RU payload designed for a tethered Helikite, and a 5G UE/small cell payload designed for a drone. The Helikite RU can provide several kilometres of coverage with Line-of-Sight (LoS), directly serving UEs or providing 5G/WiFi backhaul for the drone payloads. The drone payload can be configured as a 5G UE or a 5G small cell, to provide rapid WiFi or 5G coverage. The testbed components include:

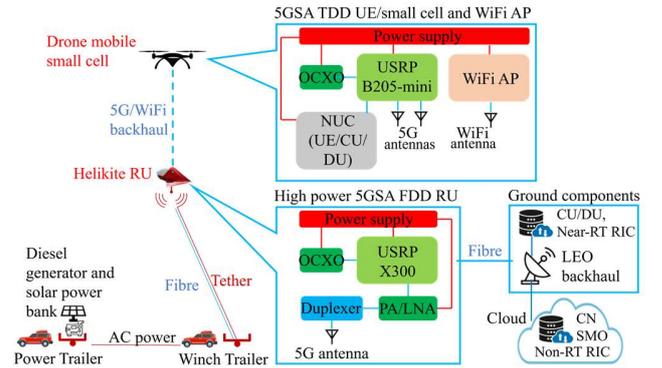

Fig. 2. Aerial network testbed design

- Ground logistics: a winch trailer which controls the launch and recovery of the Helikite and a power trailer with solar cells and a diesel generator. The winch trailer also houses a fibre drum which provides fronthaul for the Helikite RU.
- Helikite payload: a 5G RU (USRP-X300, supports frequencies below 6 GHz) which can be configured to operate Frequency-Division Duplexing (FDD) or Time-Division Duplexing (TDD) with appropriate external RF components. Fig. 2 shows an example of FDD configuration with external amplifiers, a cavity duplexer, and a single antenna. The payload also includes a 10 MHz reference clock and a WiFi bridge which provides backhaul for the drone payload.
- Drone payload: a lightweight USRP B205-mini which can be configured as a 5G UE or small cell. An onboard mini-PC (e.g., NUC) hosts the open source CU/DU/UE software (SRSRAN [14]). Similarly, the payload can be FDD or TDD (Fig. 2 shows a TDD example). A WiFi Access Point (AP)/bridge is used to either provide WiFi access for ground users (with a backhaul from 5G UE) or backhaul for the 5G small cell.
- Edge and cloud components: an on-site workstation hosting the CU/DU and Near-RT RIC software connected to the Helikite RU via a fibre fronthaul. A LEO backhaul (or any other forms of backhaul) is used to connect the edge components to the CN (Open5GS), SMO and Non-RT RIC on the cloud. The cloud components can be moved to the edge if appropriate computing resource is available.

## IV. IMPLEMENTATION DETAILS

### A. Helikite Testbed Implementation

In this subsection, we present the details of the implemented Helikite 5G RU testbed. The upper half of Fig. 3 shows the pictures of the testbed taken during one of the Helikite flights, which include the two trailers and the 21 m$^3$ Helikite (10 kg payload capacity). Multiple versions of the payload have been developed to suit different capabilities of the Helikites and to meet functionalities required by the use case. A single low-power FDD cell payload (lower left of Fig. 3)) has been developed for events requiring network coverage of a small cell. The average output power of the USRP-X300 is measured at -2 dBm. The antennas used on all payloads are omni-directional with a maximum gain of 2 dBi. All cells are single-input and single-output (SISO). The exclusive functionality provided by this payload is video streaming with a steerable GoPro camera. This payload weighs 8 kg and is suitable for fly with both the 21 $m^3$ Helikite and the 34 $m^3$ Helikite (15 kg capacity).

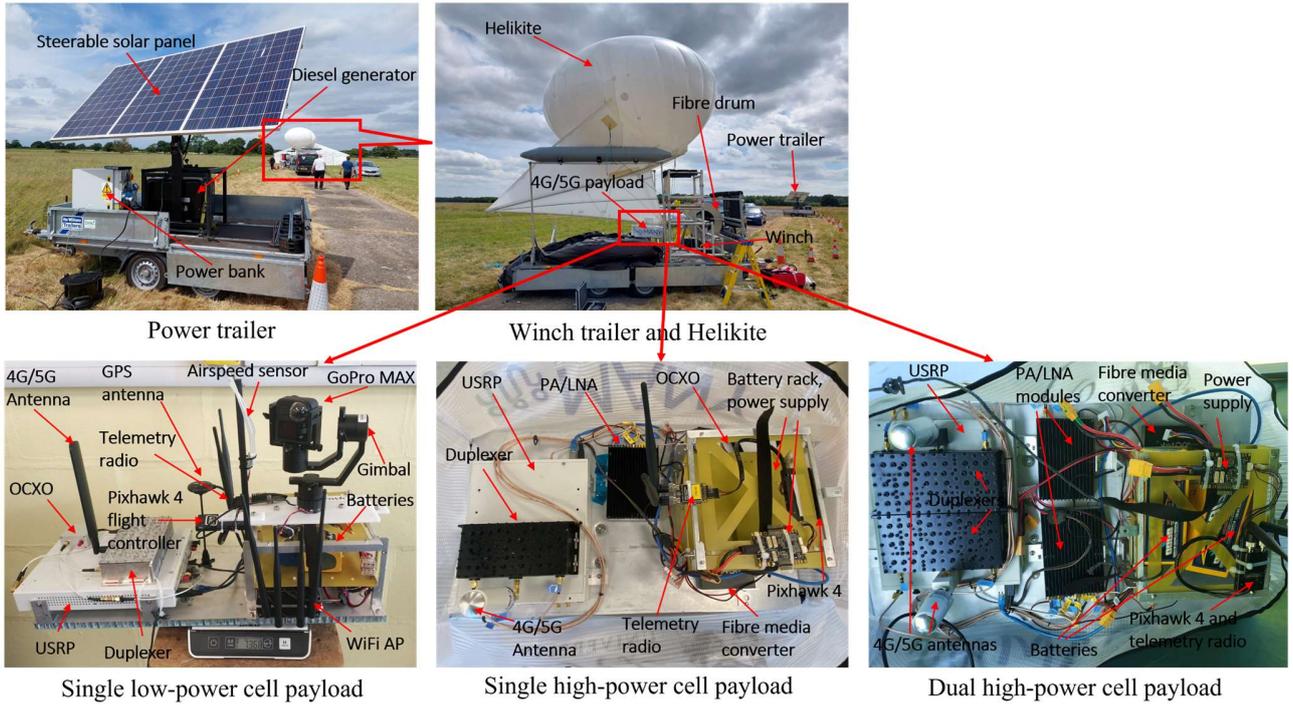

Fig. 3. System components and different versions of the 4G/5G payload

A single high-power cell payload (lower middle of Fig. 3) capable of LTE and 5GSA has been developed for events requiring larger network coverage. With external amplifiers (maximum 43 dBm output) this payload offers much better coverage than the low-power version. Detailed coverage test results will be presented in Section V. This payload weighs 9 kg and is suitable to fly with both Helikites. A dual high-power cell payload (lower right of Fig. 3) has been developed to operate an LTE cell and a 5G NSA cell to support both 4G and 5G handsets. This payload provides similar coverage as the single high-power cell payload. This payload weighs 12 kg and is only suitable to fly with the 34 $m^3$ Helikite. A similar system has been configured specifically for rapidly deployable TN usage with a slightly different packaging to make it suitable for transport in a ground vehicle.

### B. O-RAN Software Implementation

A set of open source O-RAN software is selected to support the Helikite testbed. Open5GS is used as the 5G CN, SRSRAN is selected as the CU/DU and FlexRIC is used as the NearRT-RIC. The SRSRAN CU/DU support split options 7.2 and 8, and are compatible with SDRs as well as off-the-shelf 7.2 split RUs. The support of KPM and RAN Control (RC) service models (SMs) makes performance monitoring and optimisation possible. The FlexRIC is a C binary NearRT-RIC which supports KPM, RC and several other customised E2SMs. Guidelines can be found in [15] for configuring the selected O-RAN software. A video[2] is made to showcase our selected set of O-RAN software. There are other open source options available such as the free5GC CN, OpenAirInterface CU/DU and NearRT-RIC/NonRT-RIC from the O-RAN Software Community.

### C. ML for NTN Energy Saving

To demonstrate the benefits of the RIC and its onboard intelligence, we create a simulated NTN where the cells can be turned on/off based on KPMs. The environment is a 10 km square with a coverage cell (36 dBm, 3300 MHz, always on) deployed at the centre (1 km altitude) and 9 capacity cells (28 dBm, 3600-4000 MHz, on/off) evenly distributed (60 meters altitude). The coverage cell has a 20 MHz bandwidth and consumes more energy than the capacity cell which has a 40 MHz bandwidth. 50 static UEs are randomly located with each UE having a unique active period per day and per hour traffic pattern. We use a simple Deep Q-Network (DQN) with 3 hidden layers to generate actions of turning on/off a certain capacity cell (one action per hour) based on the input cell KPMs including on/off status, number of connected UEs, throughput, energy consumption and time. The network energy efficiency (throughput divided by energy consumption) is used as the reward to train the DQN. Compared with the baseline where all cells are always on, the intelligent on/off control reduces the network's daily energy consumption by more than 40% and improves the energy efficiency by almost 100%. Although this simple simulated environment is not realistic, it still shows the massive potential benefits that the RIC offers.

---

[2]Video available at https://youtu.be/5H_DPgyQbtM

## V. Performance Evaluations

The terrestrial and Helikite networks were implemented as part of the Mobile Access North Yorkshire (MANY)[3] project which aimed at improving connectivity for rural communities. The O-RAN software was implemented as part of the YO-RAN[4] project. This testbed evaluates how such components could eventually form segments of heterogeneous neutral host O-RAN compatible networks. This section evaluates the system performance from two different aspects, including the preparation time and operating duration, along with the coverage and network throughput performance.

### A. Preparation and Maintenance Time

The preparation time required before the network can operate and the system maintenance time (mainly battery swap) are critical for the continuity of service. A preparation time of about 2 hours (team of 4) is required beforehand for the Helikite to reach its operational altitude. The time consumption for logistics can be reduced by conducting the tasks in parallel if a larger team is available. The payloads are powered by batteries which support a 2 to 3-hour operation. It takes about 20 to 30 minutes for battery swap depending on the altitude. It is possible to avoid the network down time by sending mains power or power over Ethernet (PoE) from the ground to the Helikite with the cost of payload capacity. This was not implemented due to the risk of lightning which could harm the ground crew. Another alternative is using a second Helikite to cover the network down time.

### B. Coverage and Network Throughput Performance

Multiple field tests have been conducted to evaluate the coverage of the Helikite and TN payloads. The single low-power cell payload (FDD with10 MHz bandwidth on 2650 MHz) has been tested on the Helikite (60 meters altitude) at the University of York campus. A Notice to Aviation (NOTAM) is required for altitudes above 60 meters in most locations of the UK. The RSRP logs were recorded by a OnePlus Nord handset using the Android application Network Cell Info Lite. The handset was carried by a person (on foot) at the height of about 1.5 meters. The RSRP of the handset was usually between -70 to -80 dBm while the handset was within a few meters to the antenna before the Helikite was launched. The RSRP was between -100 to -105 dBm at about 100 meters away while the Helikite was at target altitude. At about 250 meters the RSRP dropped below the sensitivity of the handset (-125 dBm or lower).

Fig. 4 shows the test results of the terrestrial dual high-power cell payload carried out at the Arkengarthdale, UK. This location (when this test was conducted) had no mobile access from any MNOs. These tests were carried out together with Swaledale Mountain Rescue (SMRT) and Safenetics who used this payload to test the push-to-talk and push-to-video mission critical applications on Android handsets. This joint activity was reported by the BBC Click[5]. The payload was configured to operate a 10 MHz FDD LTE cell (2650 MHz) and a 10 MHz FDD 5G NSA cell (1850 MHz). The ground base was set up in a public car park on the roadside. The antennas were mounted on the roof of a van with the height of 2 meters. Fig. 4 also includes estimated RSRP (green lines) using the rural macro cell propagation models in 3GPP Technical Report (TR) 38.901 with the average street width and average building height set to 5 meters.

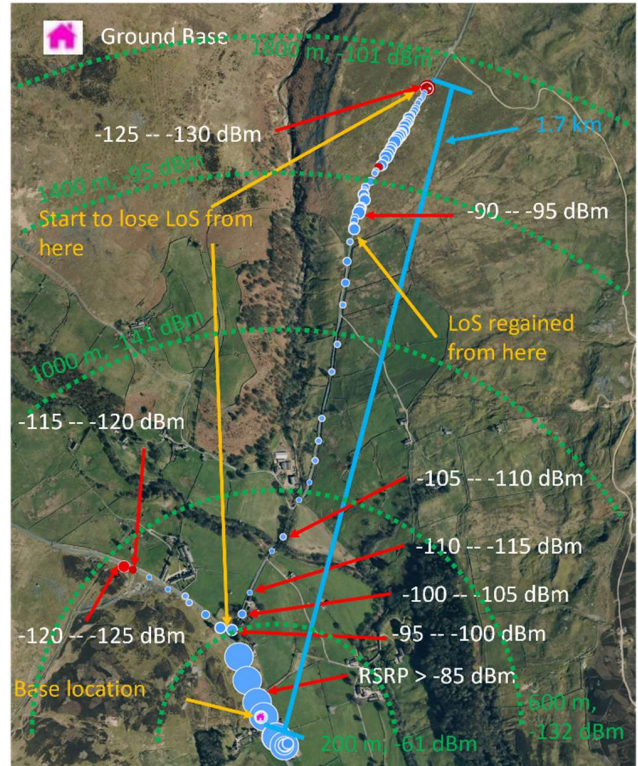

Fig. 4. Coverage of the terrestrial dual high-power cell payload at the Arkengarthdale, North Yorkshire

The handset was mostly kept inside a vehicle except the locations where the RSRP points became dense (on the road near the top of the map). At these locations the handset was carried by a person on foot. This test site was specifically chosen to have a hilly terrain with multiple levels of elevation. LoS was lost at the junction (where the paths of the logs split) where the path to the northwest blocked by a hill and the steeply downhill path to the north blocked by terrain. At about 1 km to the north of the junction the road went uphill and the LoS path to the ground base was regained. From here the RSRP was mostly kept between -90 and -95 dBm. 700 meters towards the north the road reached its highest point and caused the LoS to be blocked by the terrain again and the RSRP reduced rapidly. The experimental results matched the estimation when LoS existed and was higher without LoS. The results in Fig. 4 also indicate the difficulty in finding a LoS path in such a terrain

---

[3] DCMS MANY, available at https://mobileaccessnorthyorkshire.co.uk/
[4] DSIT YO-RAN, available at https://yo-ran.org/
[5] BBC Click – The Return of WMC, available at https://www.bbc.co.uk/iplayer/episode/m00157k2/click-the-return-of-mwc

Accepted for publication in IEEE Network Magazine, the copyright has been transferred to IEEE

(which is common in underserved rural areas) when the handset and the RU are both at ground level. The LoS coverage can be significantly improved if the RU is on a Helikite.

Fig. 5 shows the test results of a single high-power cell payload on the Helikite carried out at the Elvington Airfield to the southeast of York, UK. The runway offers continuous LoS visibility. The payload was configured to operate as an FDD LTE cell with 10 MHz bandwidth at 2630 MHz. Note that by the time this flight was conducted (June 2022), the SRSRAN 5GSA was not stable enough for flight tests, therefore, LTE was used. The reliability was improved with later SRSRAN releases, and no hardware changes were required to operate 5GSA. The 21 m³ Helikite was used and its altitude was kept between 80 and 100 meters. At this airfield location a NOTAM was not required for altitudes below 120 meters. The handset was kept inside a vehicle. Clear LoS existed while the handset was within the airfield perimeter and the RSRP at the entrance of the site (about 2 km away from the Helikite) was higher than -100 dBm. After exiting the airfield, the paths of the logs split into two directions. From this point LoS propagation became unreliable because of the tall trees with dense foliage on the roadsides and other vehicles on the road (particularly lorries). Estimated RSRP (green lines) were also included for comparison, with an additional 6 dB of shadowing because the handset was kept inside a vehicle.

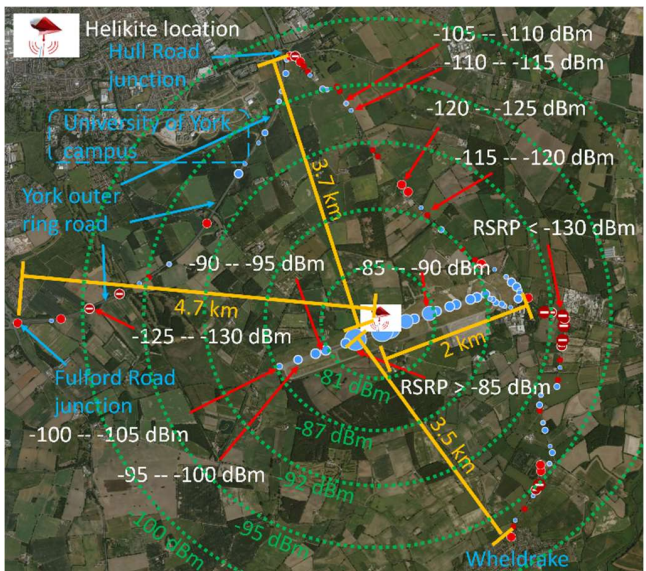

Fig. 5. Coverage of the single high-power cell on the Helikite at the Elvington Airfield

Table I shows the throughput performance of the networks provided by the payloads and the WiFi AP on the ground. The throughput of the WiFi AP (5 GHz band) was tested using the Android speedtest application by Ookla on a OnePlus Nord. The throughput of the 4G/5G networks was tested with the same handset (in an indoor lab) and an iPerf3 server at the CN.

### C. Lessons Learned and Discussions

In this subsection we discuss the advantages and disadvantages of the implemented Helikite testbed, and challenges and lessons learned during the experiments.

TABLE I
NETWORK THROUGHPUT TEST RESULTS

| Network type | Downlink (Mbps) | Uplink (Mbps) |
|---|---|---|
| WiFi (LEO satellite) | 300 | 20 |
| LTE (10 MHz bandwidth, FDD) | 46 | 21 |
| 5G NSA (10 MHz bandwidth, FDD) | 28 | 16 |
| 5G SA (20 MHz bandwidth, TDD) | 67 | 36 |
| 5G SA (100 MHz bandwidth, TDD) | 306 | 105 |

Compared with other deployment platforms, the Helikite testbed (or the tethered balloons in general) provides benefits of:

- Flexible locations and rapid temporary deployments (with the complete system on two trailers) compared with terrestrial infrastructure.
- Higher altitudes than terrestrial base stations and therefore the better LoS coverage.
- Much better operation time than drones given the nature of the helium balloon.
- Generally higher payload capacity than drones.
- The tethered feature allows the fronthaul to be delivered via fibre which is much simpler to implement than a wireless fronthaul.

The testbed also comes with drawbacks such as:

- Less capable to relocate compared with drones and HAPs.
- Smaller coverage compared with platforms at higher altitudes.

We have also experienced challenging issues during the experiments which bring difficulties in practical deployments:

- Safety of operation under windy weather. The manufacturer claims the potential to operate a Helikite with up to 50 mph wind, however, handling the Helikite near the ground (e.g., launching and retrieving) with the wind speed of above 30 mph is extremely difficult.
- Aftercare and storage. The envelope of the Helikite needs to be dried before folding for storage otherwise the residual moisture could damage the envelope and cause helium leakage.
- Availability of helium. We have occasionally experienced difficulty of helium supply due to the global shortage of helium. There are also constraints of recycling helium due to the high cost of a helium pump and its limited compressing capability.

As our experiments show, both environment and operating frequency contribute to the network performance. Terrestrial deployments suffered coverage issues caused by shadowing of the terrain which can be significantly improved by NTN. Due to the constraints of acquiring RF licenses we have conducted experiments largely on FDD bands below 2.7 GHz. The coverage difference needs to be considered while deploying TDD networks particularly on higher frequencies such as the n77 band (3.7 GHz) which is commonly used by commercial RUs. The additional attenuation can be mitigated by MIMO and beamforming.

## VI. CHALLENGES AND FUTURE PATHWAYS

Many challenges remain for implementing the proposed heterogeneous network and in this section, we discuss the challenges to be tackled and the future pathways.

- Reliable fronthaul for LAPs: our implemented Helikite RU uses a fibre fronthaul; however, this is not available for fast moving drones, and it is also the reason that our design in Fig. 2 has the CU/DU onboard. Common technologies capable of delivering high bandwidth, low latency frontal include mmWave and Free-space optical communication (FSO), both of which are difficult to implement on drones, and require rapid beamforming or mechanical tracking.
- RIC conflict management: it is very likely for xApps/rApps designed for different purposes to make contradictory decisions. For example, a traffic steering xApp may move a UE to a cell with low load which is about to be turned off by an energy saving rApp. The RIC needs to monitor the conflicted decision and prioritise actions based on the condition of the network (potentially as a specific xApp/rApp).
- RAN/RIC interoperability: unlike the all-in-one solutions provided by the telecommunication equipment vendors, the advantage of neutral host networks is that the MNOs can deploy their networks onto the neutral infrastructure according to the capacity needed. To achieve this, interoperability across RAN hardware/software components and RICs (and the hosted xApp/rApps/zApps) from different vendors is required. The NTN infrastructure brings additional challenges since the RIC may require access to platform specific interfaces for tasks such as route planning.
- RAN actuations: the support of certain RAN actuations are required for the intelligent NTN control. For example, HO is necessary for almost all aforementioned optimisation directions, and RC and Cell Configuration and Control (CCC) E2SMs are required for energy saving and traffic steering. The support of such actuations varies across vendors but is generally in early stage of development with limited capability.
- Payload design: customised payload design is not efficient considering the variety of NTN platforms particularly when they need to be rapidly deployed. Modularisation can be used to design functional RAN components for different SWaP capabilities which can be flexibly coupled to build payloads for different platforms.
- Regulations: local RF emission and aviation regulations need to be complied while deploying NTNs. For example, in the UK, an innovation and trial license can be applied from Ofcom for non-commercial use and an operational license on shared bands (largely on n77) can be applied for commercial use. The operator also needs to submit a NOTAM to the Civil Aviation Authority (CAA) to use aerial platforms within an Aerodrome Traffic Zone or for altitudes of above 60 meters.

## VII. Conclusions

This paper proposed a heterogeneous network architecture involving multiple layers of TNs and NTNs adapted to the O-RAN framework. With the support of the AI/ML models onboard the RIC, NTN specific management and optimisation targets (such as flight management and traffic steering) can be delivered. A rapidly deployable neural host network testbed has been developed to use on terrestrial vehicles and tethered Helikites to provide 4G/5G coverage without relying on existing infrastructure. The TN testbed can be ready for operation within 15 minutes deployed with the support of 1 to 2 people, and the Helikite testbed can be operational within two hours (team of 4 people). Detailed experiments have been carried out to evaluate the coverage and throughput performance of the testbed. We have also discussed the challenges for delivering the proposed heterogeneous network which indicated the future pathways for research and development.